\documentclass{elsart}
\input epsf

\usepackage{amssymb}

\begin{document}

\begin{frontmatter}



\title{The QCD Mixed Phase: Shaken, not Stirred}


\author{David Bower and Sean Gavin}

\address{Physics Department, Wayne State University, Detroit, MI 48201}

\begin{abstract}
Baryon fluctuations exceeding Poisson expectations can signal a nearly
first order phase transition at RHIC. We show how these fluctuations
can be measured, and apply a dissipative-hydrodynamic formulation used
in condensed matter physics to simulate their evolution.
\end{abstract}


\end{frontmatter}

\small
If the phase transition from quark matter to hadron gas is first
order, matter at the appropriate temperatures and densities can exist
as a mixed phase consisting of plasma droplets in equilibrium with a
surrounding hadronic fluid.
%
%
%
%
If formed in ion collisions, this mixed phase can produce large
event-by-event fluctuations as the system hadronizes. Extraordinary
baryon fluctuations \cite{Gavin} can accompany a first order
transition at high baryon density \cite{Rajagopal} and, possibly, a
near transition at zero baryon density \cite{Gavai},
\cite{McLerran}.  In ref.~\cite{Gavin}, we argued that baryon number
conservation and rapid longitudinal expansion limits the extent to
which post-hadronization interactions can erode fluctuations in
a rapidity interval. Here, we further explore the rise and fall of
these fluctuations using real-time lattice simulations \cite{Bower}.
  
At high baryon density, QCD with two massless flavors can exhibit a
first order transition whose coexistance curve culminates in a
tricritical point at temperature $T_c$ and baryon chemical potential
$\mu_c$ \cite{Rajagopal}.  For $T>T_c$ and $\mu<\mu_c$, a second order
phase transition breaks/restores chiral symmetry. If the quark masses
are sufficiently large, the second order transition is replaced by a
smooth transformation (chiral symmetry is explicitly broken). The
first order line remains, however, with the tricritical point replaced
by a critical point in the same universality class as a liquid--gas
transition.
  
At RHIC, baryon density may also serve as an approximate order
parameter for the nearly first order transition at small net baryon
density. Lattice simulations \cite{Gavai} and general arguments
\cite{McLerran} show that the baryon susceptibility $\chi$ at $\mu = 0$ 
can increase suddenly as temperature is increased near $T_h\sim
150$~MeV, where the chiral order parameter and the energy density
change sharply.  Jumps in the susceptibility commonly accompany first
order transitions.  For a liquid-gas transition, $\chi =
\partial\rho/\partial \mu$ is proportional to the compressibility:
steam is much more compressible than water.
  
Large fluctuations in baryon number occur during phase separation in a
first order transition. Figure 1 (left) shows the phase diagram in the
$T-\rho$ plane, where $\rho$ is the baryon density. A uniform system
quenched into the outer parabolic region below $T_c$ will separate into droplets at
high baryon density $\rho_q$ surrounded by matter at density $\rho_h$. The
net baryon number $N_B$ in a sub-volume of the system varies depending
on the number of droplets in the sub-volume. The variance of the
baryon number $V= \langle N_B^2\rangle - \langle N_B\rangle^2$ can
exceed the equilibrium expectation by an amount
\begin{equation}\Delta V \approx f(1-f)(\Delta N_B)^2,
\label{eq:fluct}\end{equation}
\vskip -0.2in
where $f$ is the fraction of the high density phase in the
sub-volume $V$ and $\Delta N_B = (\rho_q - \rho_h)V$. 
We will argue that nonequilibrium evolution in ion collisions can
allow these fluctuations to survive post-hadronization evolution.
\begin{figure} 
\epsfxsize=5.25in
\leftline{\epsffile{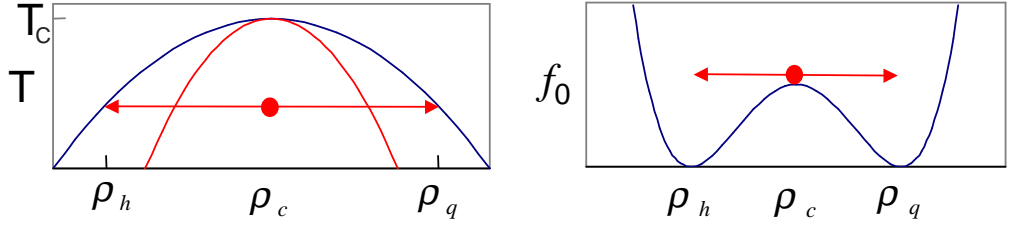}}
\caption[]{
  Phase diagram (left) and free energy (right) vs. baryon density for (3).
}\end{figure}

We stress that a super-poissonian variance such as (\ref{eq:fluct}) is
straightforward to test experimentally by measuring
\begin{equation}
\Omega_p = (V_p-\langle N_p\rangle)/\langle N_p\rangle^2,
\label{eq:Omega}\end{equation}
\vskip -0.2in
where $N_p$ is the number of protons in a rapidity interval and $V_p$
is its variance.  This quantity vanishes in equilibrium and is related
to the more familiar scaled variance $\omega_p = \langle
N_p\rangle(1+\Omega_p)$. Significantly, we find that $\Omega_p$ equals the
total $\Omega_B$ -- which includes unseen neutrons -- for a range of thermal and
Glauber models that respect isospin symmetry.  Specifically, $\Omega_p
= \Omega_B$ because the probability of $N_p$ satisfies $p(N_p) = \sum_{N_B}
p(N_B)p(N_B|N_p)$ with a binomial distribution $p(N_B|N_p)$ for $N_p$ at
fixed baryon number $N_B$.
Isospin fluctuations can alter $p(N_B|N_p)$ near the tricritical point
or in the presence of a disoriented chiral condensate, but that will
be evident from pion measurements. 

To describe the evolution of
the inhomogeneous mixed phase, we follow the standard condensed matter
practice and write a Ginzburg-Landau free energy:
\begin{equation}
f=\kappa(\nabla \rho)^2/2 + f_0, \,\,\,\,\,\, \,\,\,\,\,\, 
f_0 = -m^2(\rho-\rho_c)^2/2 + \lambda (\rho-\rho_c)^4/4
\label{eq:phi4}\end{equation}
\vskip -0.2in
where $f_0(\rho)$ describes the excursions of the baryon density
$\rho$ from its equilibrium value in the uniform matter. The $\kappa$
term describes the droplet surface tension, $\sigma\propto
\kappa^{1/2}$. For $m^2\propto T_c-T$ we find the correct liquid-gas
critical exponents. The values $\rho_h$ and $\rho_q$ in fig. 1 (right)
correspond to the equilibrium densities at $T < T_c$. To
describe the dynamics of the system, we must account for the fact that
baryon number is conserved. Furthermore, it is crucial to include
dissipation to describe this strongly fluctuating system. The simplest
equations that meet these criteria are:
\begin{equation}
\partial \rho/\partial t = M\nabla^2\mu, \,\,\,\,\,\, \,\,\,\,\,\, 
\mu = f_0^\prime -\kappa\nabla^2\rho;
\label{eq:diff}\end{equation}
\vskip -0.2in
model B in \cite{Bray}. We identify $D=2m^2M$ as the baryon
diffusion coefficient by linearizing (\ref{eq:diff}) about $\rho_{h}$.
  
To describe nuclear collisions, we
extend (\ref{eq:diff}) to include drift due to Bjorken longitudinal flow:
\begin{equation}
\partial \rho/\partial \tau + \rho/\tau = M\nabla^2\mu,
\label{eq:drift}\end{equation}
\vskip -0.2in
where $\tau$ is the proper time and $\mu$ is given by
(\ref{eq:phi4}, \ref{eq:diff}). The new term forces the average density to decrease
as $\langle\rho\rangle\propto \tau^{-1}$, driving the system through
the phase coexistence region. Fluctuations grow when densities are
near $\rho_c$ (c.f. fig.~1). 
\begin{figure} 
\epsfxsize=5.25in
\centerline{\epsffile{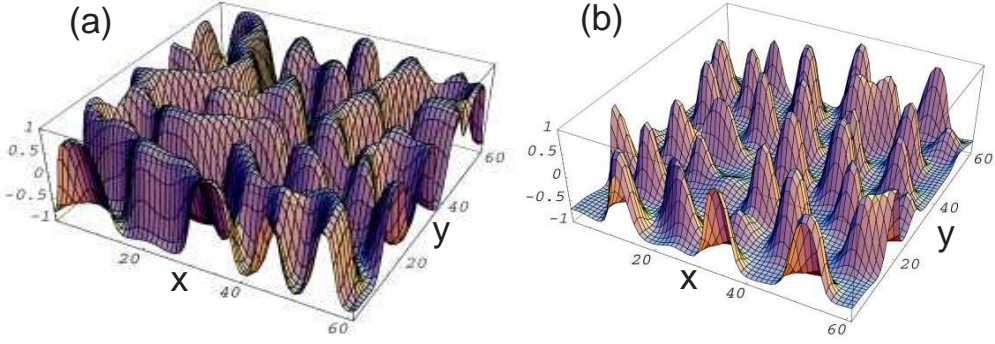}}
\caption[]{
  Order parameter in the transverse plane without (a) and with expansion (b).
}\end{figure}

Phase separation is most dramatic if the rapid expansion drives the
system into the unstable region; i.e., the
inner parabolic region in fig.~1 (left), corresponding to 
$f_0^{\prime\prime}(\rho) < 0$. Droplets form from runaway density
fluctuations in a process known as spinodal decomposition. Linearizing
near $\rho_c$, we estimate the time scale for this process to be
$\tau_R = 8\xi^2/D$, where $\xi =\kappa^{1/2}/m$ is the correlation
length.  For times $t\gg \tau_R$ the system undergoes a nonlinear
evolution in which droplets merge, reducing their surface energy.
  
Figure 2 compares 2+1 dimensional numerical simulations of
(\ref{eq:diff}) and (\ref{eq:drift}) for $\psi = (\rho
-\rho_c)/\rho_c$ in the transverse plane.  We take $\tau_R$ and $\xi$
to each be 1~fm as motivated in \cite{Gavin}. ($D\sim 8$~fm is
consistent with calculations in \cite{Prakash}). The
expanding system reaches $\rho_c$ at $\tau_0 = 5$~fm.  Expansion 
shown in fig.~2b prevents droplets from merging as in fig.~2a. Because
this is a dissipative system, we must apply thermal noise at each
lattice site at $\tau_0$ to seed phase separation. The memory of the
initial conditions is essentially lost for $\tau - \tau_0 > \tau_R$.
  
Figure 3 shows the computed variance for two different initial times
and for two rapidity intervals. The variance is computed from a sample
of 5000 simulated events, each unique due to the thermal noise. We
see that the super-poissonian fluctuations grow appreciably by $\tau
\sim 2\tau_0$. This variance drops as the rapidity interval is
increased. We find that variance is governed by the ratio
$\tau_0/\tau_R$, which compares the expansion and droplet-growth time
scales. 
  
We emphasize that these calculations include diffusion, which dampens
the fluctuations once the system becomes stable. For (\ref{eq:phi4})
with $\langle \rho\rangle \propto \tau^{-1}$, the system is unstable
only for $\tau < 2.3\;\tau_0$.  We extend the calculations to much
longer times to demonstrate that the fluctuations in rapidity survive
well past the freezout time, of order 10--30 fm, in accord with
\cite{Gavin}. We comment that convection, viscosity and
collision-geometry effects can reduce $\Omega_p$ compared to
fig.~3. Moreover, our phase transition effect may be compensated to
some extent by the effect discussed by Koch and Asakawa in these
proceedings, which owes to the difference between fluctuations in a
plasma compared to a hadron gas. Nevertheless, the strength of the
signal in our exploratory calculations invites further work.
  
\begin{figure} 
\epsfxsize=3.25in
\centerline{\epsffile{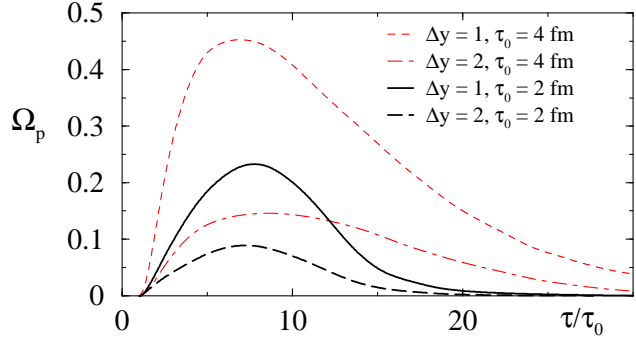}}
\caption[]{
  Enhanced variance vs. time for two rapidity windows, eq~(1).
}\end{figure}
We thank R. Bellwied, P. Braun Munzinger, K. Elder, M. Grant,
J. Kapusta, G. Kunde, P. Keyes, B. M\"uller, I. Mishustin, C. Pruneau and
S. Voloshin.

\end{document}